\documentclass[prl,amsmath,amssymb,superscriptaddress,twocolumn]{revtex4}
\usepackage{graphicx}

\newcommand{\br}{{\bf r}}
\newcommand{\bk}{{\bf k}}

\newcommand{\bq}{{\bf q}}
\newcommand{\bA}{{\bf A}}

\newcommand{\ms}{{m^{\ast}}}

\newcommand{\eps}{\epsilon}

\DeclareMathAlphabet{\mathpzc}{OT1}{pzc}{m}{it} 
\begin{document}
\title{Disorder induced non-Fermi liquid near a metal-superconductor quantum phase transition.}
\author{Oskar Vafek}
\affiliation{Department of Physics, Stanford University, CA 94305}
\author{M. R. Beasley}
\affiliation{Department of Applied Physics, Stanford University,
Stanford, CA 94305}
\author{Steven A. Kivelson}
\affiliation{Department of Physics, Stanford University, CA 94305}
\date{\today}
\begin{abstract}
We study the flux-driven superconductor-metal  transition in
ultrasmall cylinders observed experimentally by Liu {\em
et.al.}\cite{Liu01}. Where $T_c\to 0$, there is a quantum critical
point, and a large fluctuation conductivity is observed in the
proximate metallic phase over a wide range of $T$ and flux .
However, we find that the predicted (Gaussian) fluctuation
conductivity in the neighborhood of the quantum critical point is 4
orders of magnitude smaller than observed experimentally. We argue
that the breakdown of Anderson's theorem at any non-integer flux
leads to a broad fluctuation region reflecting the existence of
``rare regions'' with local superconducting order. We calculate the
leading order correction to the conductivity within a simple model
of statistically induced Josephson coupled local ordered regions
that rationalizes the existing data.
\end{abstract} \maketitle

The physics of the zero temperature  ``superconductor-insulator"  transition in
low dimensional systems is important in its own right, and as a paradigm for quantum phase transitions in fermionic systems\cite{sondhi}.  It is therefore embarrassing that it remains an unsolved and contravertial problem. Most studies have focussed on cases in which the  transition is either driven by disorder or a
combination of disorder and magnetic field.  However, these important studies often reveal the existence of anomalous metallic (dissipative) phases where superconductivity is destroyed, rather than the expected insulating phases. (Some representative reviews of this subject are contained in Refs. \cite{sondhi,liu2,mason,ppdd}.)

Moreover, it is presently unclear whether the range of parameters
({\it e.g.} temperature, $T$, and magnetic field) over which quantum
critical fluctuations dominate the behavior of such systems is large
or small -- this later uncertainty likely has more general
implications for the issue of whether or not quantum critical
fluctuations underly a host of ``non-Fermi liquid'' properties seen
in highly correlated materials from heavy fermion metals to the high
temperature superconductors.  (See, for example, Refs.
\cite{varma,coleman,chakravarty,sachdev}.)

We study here a concrete and particularly simple version of this
problem that was recently studied experimentally by Liu {\em
et.al.}\cite{Liu01}:  As shown in the inset of Fig. 1, the system
consists of a thin insulating cylinder coated with a simple BCS
metal. Magnetic flux of magnitude $\phi$ is threaded along the
central axis. The quantum phase transition is driven by the balance
between the kinetic energy stored in the flux induced diamagnetic
currents and the condensation energy, i.e. this represents the
extreme limit of the Little-Parks experiment. Experimentally, it was
observed that at a critical value of  $\phi=\phi_c$ there is a
continuous quantum phase transition from a superconductor to a
metallic phase, where $\phi_c < \Phi_0/2$ and $\Phi_0=hc/2e$ is the
superconducting flux quantum. However, even well beyond the critical
flux, where $\phi\approx \Phi_0/2$, the $T=0$ resistance is
$\sim0.3$ of its normal state value, indicating the persistence of
substantial superconducting fluctuations.

We have found that the value of $\phi_c$ can be obtained with
surprising accuracy from simple mean-field theory. However,
standard\cite{AL} small amplitude (Gaussian) fluctuations about the
mean-field solution are at least four orders of magnitude too small
to account for the observed fluctuation conductivity. Rather, we
argue that a breakdown of Anderson's theorem (due to the breaking of
time-reversal symmetry by the applied flux) combined with the
physics of rare regions characteristic of disordered systems (akin
to the physics of Lifshitz tails in semiconductors) provide a
natural explanation for at least the bulk of the observed
fluctuation effects. (The nature of the non-Fermi liquid groundstate
at $T=0$ remains an intriguing, but still unsolved problem.)

\begin{figure}
\includegraphics[width=0.4\textwidth]{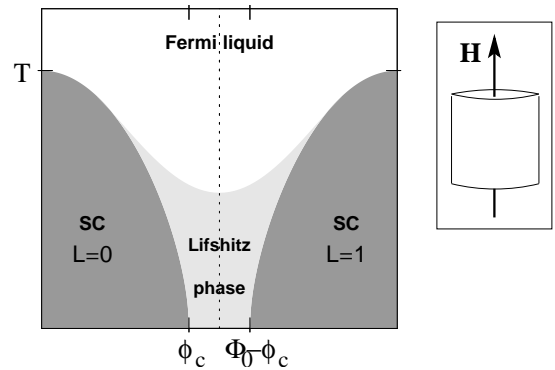}
\caption{Schematic phase diagram for the flux induced quantum phase
transition in a thin walled superconducting cylinder. The
superconductivity survives inside each dome, while at temperatures
above the mean-field $T_c$, the system is a Fermi liquid. The
``Lifshitz phase'' at intermediate $\phi$ is dominated by
superconducting fluctuations in rare regions with a larger than
average local $T_c$.  The dotted line denotes half the flux quantum
$\Phi_0/2=hc/4e$. Inset - sketch of the experimental geometry.}
\end{figure}

For $\phi=0$, the system is time reversal invariant, hence
Anderson's theorem implies that $T_c(\ell,\phi=0)$ is independent of
the mean-free path, $\ell$, over a broad range of $\ell > k_F^{-1}$,
where  $k_F$ is the Fermi wave number.  Hence, although there are
always mesoscopic variations in the local values of $\ell$, $T_c$
remains uniform, and indeed the observed resistive transition is
very sharp.

On the other hand, at non-zero $\phi$, time reversal is no longer a
symmetry, and hence $T_c(\ell,\phi)$ varies with $\ell$;
correspondingly, at $T=0$, the critical value of the flux is (as
shown in Eq. \ref{Tcell}) a strong function of $\ell$ (For related
issues see Refs. \cite{Larkin81,galitski01,Millis01}). As a
consequence, even when $\phi$ is greater than the average value of $
\phi_c$, local statistical fluctuations of $\ell$ lead to rare
regions with a non-zero local $T_c$.  At not too low temperatures,
the effect of these regions on the metallic conductivity can be
calculated in a perturbative expansion in the Josephson coupling
between them, $E_J$ (See Eq. \ref{resist}.) The phase dynamics are
dominated by their coupling to the dissipative environment, and as a
result they can make a substantial contribution to the conductivity
even when $T>E_J$. The result is a broad region of the $\phi-T$
plane, which we have called the ``Lifshitz-phase,'' in which the
resistance is reduced from its value in the Fermi liquid phase.

The Hamiltonian for a single electron confined to the cylinder or
radius $R_0$ is
\begin{equation}
\mathcal{H}=\frac{1}{2\ms}\left[\frac{\hbar}{i}\nabla-\frac{e}{c}\bA\right]^2+V_{imp}(r)
\end{equation}
where we work in a gauge in which $\bA$ points in the
azimuthal direction with $A_{\theta}=\phi/(2\pi r)\approx
\phi/(2\pi R_0)$. Since most of the  flux threads through the insulating
midsection, the magnetic field is small despite the large value of the flux, so the Zeeman coupling is negligible. As long as
the magnetic length is longer that the thickness of the thin film
coating, any additional orbital effects associated with the
curvature of the electronic paths can be ignored as well.

The disorder averaged pairing susceptibility of a degenerate
assembly of such electrons is given by \cite{deGennes66}
\begin{equation}
\chi_p(\bk,i\Omega)=N(0)k_BT \sum_{\omega_n}\int d\xi d\xi'
\frac{g(\bq,\xi-\xi')}{\left(\xi-i\omega_n\right)\left(\xi'+i\omega_n-i\Omega\right)}
\nonumber
\end{equation}
where $\bq=\zeta^{-1}-\bk$, $\zeta=(\Phi_0/\phi)R_0$, and
$g(\bq,\lambda)\approx\frac{1}{\pi}\frac{D\bq^2}{\lambda^2+D^2\bq^4}$
for diffusive electrons with the diffusion constant $D=v_F \ell/3$.
The above integrals can be evaluated, yielding
\begin{equation}
\chi_p(\bk,i\Omega)=N(0)\int_0^{\omega_D}\!\!\!\!d\eps
\tanh\left[\frac{\eps}{2T}\right]\frac{4\eps}{4\eps^2+(D\bq^2+|\Omega|)^2}.
\nonumber
\end{equation}
As usual, in the weak coupling limit the condition $1=g\chi_p(0,0)$
determines the critical temperature. Near the quantum critical
point, $\chi_p(\bk,i\Omega)$ can be expanded in powers of $T^2$ and
we get
\begin{eqnarray}
\chi_p(\bk,i\Omega)=\frac{1}{2}N(0)
\left(\ln\left[1+\left(\frac{2\omega_D}{D\bq^2+|\Omega|}\right)^2\right]\nonumber
\right.\\
\left. -\frac{4\pi^2}{3} \frac{T^2}{(D\bq^2+|\Omega|)^2}
\right)
\end{eqnarray}
which means that near the quantum critical point
\begin{equation}
T_c(\phi)\sim T_c(0) \sqrt{\frac{\phi_c-|\phi-n\Phi_0|} {\phi_c}} \
\ {\rm for} \ \ |\phi-n\Phi_0| < \phi_c \label{Tcell}
\end{equation}
%with $\nu=1/2$,
where
\begin{equation}
\phi_c=\Phi_0\sqrt{{3}/{\pi}} R_0[\ell\xi_0]^{-1/2}. \label{phic}
\end{equation}
%$\phi^{(0)}_c$
gives the width of the superconducting domes in Fig. 1 centered
around each integer flux $\phi=n\Phi_0$, so long as  $\phi_c <
\Phi_0/2$; otherwise, $T_c$ never vanishes, but rather reaches a
minimum, non-zero value for $\phi=\Phi_0/2$.

For $R_0=60\mbox{nm}$, the
superconducting coherence length in clean Al $\xi_0=1600\mbox{nm}$,
and the mean-free path, derived from the normal state conductivity,
$\ell=12\mbox{nm}$, Eq. (\ref{phic}) yeilds a theoretical
estimate, $\phi_c= 0.42\Phi_0$, which agrees with experiment 
\cite{Liu01} to better than 5\% accuracy.
The agreement with the experiment, in addition to the narrowness of the 
transition
regime in the absence of flux \cite{Liu01}, demonstrate that the
superconductivity, as assumed, has a bulk character and the finite size of
the cylinder radius only serves to determine $\phi_c$.

Inspired by the success of the mean field theory for $\phi_c$, we
now turn to the question of the character of the disordered state
when $\Phi_0/2\ge\phi>\phi_c$. In particular, we calculate the extra
conductivity on the disordered side. The action governing the
dynamics of the fluctuating order parameter is
\begin{equation}\label{clean}
S=\int
\left[K_{\tau-\tau'}(\br-\br')\Delta^{\dagger}(\tau,\br)\Delta(\tau',\br')+
b_0 |\Delta(\tau,\br)|^4\right]
\end{equation}
where $\Delta$ is the charge $2e$ pairing field, and the integral
ranges over space-time; $K(\omega,\bq)=g^{-1}-\chi_p(\omega,\bq)$
and
\begin{eqnarray}
b_0=-\frac{1}{2}\frac{N(0)}{(4\pi T)^2}\left[(1-2(\ell
q)^2)\psi^{(2)}\left(\frac{1}{2}+\frac{D\bq^2}{4\pi T}\right)+\nonumber\right. \\
\left.(1+(\ell q)^2)\frac{D\bq^2}{4\pi T}
\frac{1}{3}\psi^{(3)}\left(\frac{1}{2}+\frac{D\bq^2}{4\pi
T}\right)\right]
\end{eqnarray}
where $\psi^{(n-1)}$ is the $n-$th logarithmic derivative of the
Gamma function.

Within the Aslamazov-Larkin theory fluctuation conductivity in
d-dimensions is
\begin{equation}
\Delta\sigma_{AL}=\frac{4e^2}{h}\frac{\Gamma(2-d/2)}{(4\pi)^{d/2}}
\frac{\zeta^{4-d}}{\hbar D}\frac{k_BT}{m^{4-d}}
\end{equation}
where the mass $m$ is given in Eq. (\ref{mass}). This contribution
vanishes at $T=0$ except possibly at the quantum critical point.
Additional effects coming from Maki-Thompson and density of states
diagrams were computed by Lopatin {\em et.al.} \cite{Lopatin05}. In
addition, Ref.\cite{Lopatin05} included frequency dependence of the
boson vertex which gives a contribution at T=0, but contrary to the
experiment \cite{Liu01} leads to a {\em weak negative
magnetoresistance}. Overall, the critical regime where the Gaussian
fluctuations contribute appreciably is at least 4-orders of
magnitude smaller than observed experimentally \cite{Liu01}.

In fact, in the absence of the flux, the superconducting transition
is extremely narrow, consistent with the small value of the Ginzburg
parameter of a low T$_c$ material, while near the quantum critical
point the transition is very broad. As mentioned previously, at
non-zero $\phi$, $T_c$ is a strong function of both $\ell$ and the
cylinder radius. Since at the mean field quantum phase transition,
the Harris criterion is not satisfied, {\it i.e.} $\nu d < 2$,  one
expects a disorder driven instability of the clean critical point.
Thus, the ensemble fluctuations of $\ell$ and $R_0$ in the system
lead to randomness in the mass of the effective quantum Ginzburg
Landau theory. As a consequence, rare regions in space where
$\phi<\phi_c$ develop local superconducting order and can contribute
anomalously to the ``normal" state conductivity.

To be explicit, we model the disorder of the critical coupling
constant by a set of $N_{imp}$ delta function ``impurities" of
strength $-|u|$ positioned at random throughout the sample. This
model, while highly simplified, allows analytic treatment in the
limit of small impurity concentration, i.e. the regime marked by the
onset of the anomalous contribution to the conductivity.

Since $w$, the thickness of the Al thin film coating, as well as the
cylinder radius, $R_0$, are much less than the coherence length, the
order parameter can be taken to be a function of the position along
the cylinder, $z$, only. The resulting Ginzburg-Landau functional is
then
\begin{widetext}
\begin{equation}\label{gl}
S=2\pi R_0 w \times N(0)\int_0^{\beta} d\tau \int dz \left[\int
d\tau' G^{-1}_{\tau-\tau'}\Delta^{\dagger}(\tau',z)\Delta(\tau,z)
+\zeta^2\left|\frac{d\Delta}{dz}\right|^2+
\left(m^2-|u|\sum_{j=1}^{N_{imp}}\delta(z-z_i)\right)\left|\Delta\right|^2
+b\left|\Delta\right|^4 \right]
\end{equation}
\end{widetext}
where $\zeta=(\Phi_0/\phi)R_0$, the Fourier transform of the inverse
propagator $G^{-1}(\omega)=\zeta^2 |\omega|/D$, the contact
repulsion $b=b_0/N(0)$, and the average "mass"
\begin{eqnarray}\label{mass}
m^2&=&2\left(\frac{\phi-\phi_c}{\phi_c}+\frac{\pi^2}{3}
\frac{k^2_B}{\hbar^2}\frac{T^2\zeta^4}{D^2}\right).
\end{eqnarray}

The saddle point of the static action with a single ``impurity"  at
the origin is
\begin{equation}\label{single}
\Delta_{imp}(z)=\frac{m}{\sqrt{b}}\;\frac{1}{\sinh\left(m|z|/\zeta+\lambda\right)}
\end{equation}
where $\lambda=\tanh^{-1}\left[2m\zeta/|u|\right]$. The order
parameter takes its maximum value,
$\Delta_0=\sqrt{(u/2\zeta)^2-m^2}/\sqrt{b}$, at the origin. Note,
that this solution minimizes the action only for $2m\zeta<|u|$,
where $\Delta$ is real.

If the concentration of such impurities, $N_{imp}/L\ll m/\zeta$ ($L$
is the length of the cylinder), is small we can adopt as an ansatz
for the saddle-point solution
\begin{equation}\label{stationary}
\Delta(\tau,z)=\sum_{j=1}^{N_{imp}} e^{i\theta_j(\tau)}\Delta_{imp}(z-z_j).
\end{equation}

Substituting the above equation into a Bogoliubov-de~Gennes equation
and integrating out the fermions to leading order in the Josephson
coupling strength, $E_J$, gives an effective action for the phase
differences $\Theta_i\equiv \theta_{i+1}-\theta_i$:
\begin{eqnarray}\label{phases}
S_{eff}&=&S_0-\int_0^{\beta} d\tau
\sum^N_{j=1}E_J(j)\cos\Theta_j(\tau)
 \nonumber \\
&+&\frac{1}{\beta}\sum_{\omega_n} \sum_{j=1}^N
\alpha|\omega_n|\Theta_j(-\omega_n)\Theta_j(\omega_n)
\end{eqnarray}
where $\alpha=R_Q/R_N$, the quantum of resistance
$R_Q=h/4e^2=6.5\mbox{k}\Omega$, $S_0$ is the part of the action
independent of the phase configuration \cite{s0}, and $E_J(j)$ is
the Josephson coupling between the sites. In the dilute impurity
limit and at finite temperature, $E_J(j)$ is given approximately by
\begin{equation}
E_J(i)=\frac{2|u|}{\zeta}\frac{\Delta_{0}}{\eps_c}\;\Delta_{imp}(z_{i+1}-z_i),
\end{equation}
where $\eps^{-1}_c=2\pi R_0 w \zeta N(0)$, {\it i.e.} the Josephson
coupling falls off exponentially with $m|z_{i+1}-z_i|/\zeta$.

So long as $T$ is not too low, the correction to resistivity can be computed in powers of
 $E_J$. The most direct way is to use the Keldysh
technique \cite{FisherMPA87}, with the appropriate Keldysh
propagator matrix \cite{Kamenev04}, $G_{\mu\nu}(\omega,r)$,
$G_{12}(\omega,r)=G_{21}(-\omega,r)=G^R(\omega,r)$,
$G_{11}(\omega,r)=\coth\left[\omega/2T\right]\left(G_{12}(\omega)-G_{21}(\omega)\right)$,
and the retarded propagator is $G^R(\omega,r)=1/(i\alpha\omega)$,
which is regularized in the UV by introducing a cuttoff
$T_{UV}\approx \hbar/(k_B\tau)$. For a given disorder configuration,
the resulting correction to the normal state resistivity is
\begin{equation}
\frac{R(T)}{R_N}=1-g(\alpha)T^{-2}
\left[\frac{T}{T_{UV}}\right]^{\frac{2}{\alpha}}\frac{1}{N_{imp}}\sum_j
E_J^2(j)+\mathcal{O}(E^4_J)
\end{equation}
where $g(\alpha)=\cos[\frac{\pi}{\alpha}]
\left(\frac{\pi}{e^{\gamma_E}}\right)^{2/\alpha}\Gamma[\frac{1}{2}-\frac{1}{\alpha}]
\Gamma[\frac{1}{\alpha}]/(2\sqrt{\pi}\alpha)$. In the thermodynamic
limit, the arithmetic mean of $E^2_J(j)$ equals its (disorder)
ensemble average,
\begin{equation}
\frac{1}{N_{imp}}\sum_j E_J^2(j)=\int_0^L \!\!\!dl\; P(l) E^2_J[l],
\end{equation}
where $P(l)$ is the distribution of the junction lengths and
$E_J[l]$ is the Josephson coupling across a junction of length $l$.
In the model of disorder that we used, $P(l)$ is the probability
density of finding a variable with a value $l$ among $N_{imp}$
independent variables, each evenly distributed between $[0,L]$, and
whose sum is constrained to be $L$. In the limit of large $L$, with
the density $n_{imp}=N_{imp}/L$ held fixed, $P(l)$ can be shown to
be given by the probability distribution for the occurence of the
first success in a Poisson process, i.e. $P(l)=n_{imp} e^{-n_{imp}
l}$. Thus,
\begin{equation}
\int_0^L \!\!\!dl P(l) E^2_J[l]= \frac{\Delta^4_0}{\eps^2_c}
\left(\frac{2|u|}{\zeta}\right)^2\frac{
n_{imp}\zeta}{|u|/2\zeta+m}.\nonumber
\end{equation}
So we arrive at one of the main results of our analysis (to
$+\mathcal{O}(E^2_J)$)
\begin{equation}\label{resist}
\frac{R(T)}{R_N}=1-c\;g(\alpha)
\left[\frac{\Delta_0}{\eps_c}\right]^2\left[\frac{8|u|/\zeta}{1+2m\zeta/|u|}\right]
\frac{\Delta_0^2}{T^2}
\left[\frac{T}{T_{UV}}\right]^{\frac{2}{\alpha}}.
\end{equation} In a good metal, $\alpha\gg1$ and the above expression
has only a weak dependence on the high energy cuttoff, $T_{UV}$. The
main point is the explicit functional dependence on the scale of the
local ordering $\Delta_0$ and the density of the locally ordered
regions expressed through the dimensionless parameter
$c=n_{imp}\zeta$.

Note that in general both $\Delta_0$ and $c$ have implicit
temperature dependence. If the deviations $\delta m(z)$ from the
mean mass $m$ are $\delta-$function correlated gaussian random
variables, $\langle\delta m(x)\delta m(y)
\rangle=\frac{1}{2}W\delta(x-y)$, then for $m\gg W$, $n$, the
density of Ginzburg-Landau "impurities" (see Eq. \ref{gl}) is given
by the cumulative density of states in the Lifshitz tail below the
energy $m^2$. In 1 dimension this is given by
$n(m^2)=\frac{1}{2\pi}\frac{m}{\zeta}\;e^{-16\zeta m^3/3W}$
\cite{Halperin65}. Similarly, the strength of the impurity $u$, is
chosen by matching the average negative eigenvalue in the Lifshitz
tail below $m^2$ with the eigenenergy of the impurity bound state.
Thus $(u/(2\zeta))^2=m^2+W/(8m\zeta)$ which gives
$\Delta^2_0=W/(8m\zeta b)$.

Fitting Eq. (\ref{resist}) to the data of Ref. \cite{Liu01} yields a
qualitatively good fit for physically reasonable parameters. Hence
our theory provides a natural explanation for the broadness of the
transition observed in the experiment. Quantitative details will
require further investigation.

We conclude with a discussion of the validity of our approach. While
the dynamics of the local ordered regions can almost certainly be
described by the Eq. (\ref{phases}) at high temperature, as the
temperature is lowered, the action (\ref{phases}) must be modified.
There are several reasons: first as the temperature is lowered, the
locally ordered regions induce suppression of the single particle
density of states at the Fermi level, thereby reducing dissipation
of the collective phase variable and suppressing phase ordering. In
addition, the simple harmonic action describing dissipation is
suspect at low temperature. Finally, the quantum mechanical
tunneling of the order parameter amplitude can lead to a destruction
of the phase ordering at low temperature, giving rise to a
dissipative state\cite{spivak01}.

Acknowledgements:   We wish to acknowledge discussions with
Professors Y. Liu and B. Spivak.
O.V. was supported by the Stanford
ITP fellowship, M.R. and S.K. were supported in part by the NSF
grants DMR-0406339 and DMR-0421960 respectively.
\bibliography{bibliography}

\begin{thebibliography}{20}
\bibitem[{s0()}]{s0}
\bibinfo{note}{The self action $S_0=\frac{N_{imp}}{T}\times
  \frac{|u|}{\epsilon_c}\frac{m^2}{b} [1-\frac{2}{3}\frac{2m \zeta}{|u|}-
  \frac{1}{3}\frac{u^2}{(2m\zeta)^2}] +\delta S_0$ $
  \gamma=\epsilon_c^{-1}\frac{|u|-2m\zeta}{b} $ and $\epsilon_c^{-1}=2\pi R_0 w
  N(0)$. The first term in $S_0$ is the on-site action which is negative
  semidefinite for $2m\zeta \leq |u|$, while $\delta S_0$ is a correction from
  the small off-site overlaps.}
\end{thebibliography}
\end{document}